# Performance Issues of Heterogeneous Hadoop Clusters in Cloud Computing

[1]B.Thirumala Rao, [2]N.V.Sridevi, [3]V.Krishna Reddy, [4]L.S.S.Reddy
Department of Computer Science and Engineering,
Lakireddy Bali Reddy College of Engineering, Mylavaram
thirumail@yahoo.com, harshi.5807@gmail.com, krishna4474@gmail.com, director@lbrce.ac.in

## ABSTRACT

*Nowadays most of the cloud applications process large amount of data to provide the desired results. Data volumes to be processed by cloud applications are growing much faster than computing power. This growth demands new strategies for processing and analyzing information. Dealing with large data volumes requires two things: 1) Inexpensive, reliable storage 2) New tools for analyzing unstructured and structured data. Hadoop is a powerful open source software platform that addresses both of these problems. The current Hadoop implementation assumes that computing nodes in a cluster are homogeneous in nature. Hadoop lacks performance in heterogeneous clusters where the nodes have different computing capacity. In this paper we address the issues that affect the performance of hadoop in heterogeneous clusters and also provided some guidelines on how to overcome these bottlenecks.*

## I. INTRODUCTION

Cloud computing[1] is a relatively new way of referring to the use of shared computing resources, and it is an alternative to having local servers handle applications. Cloud computing groups together large numbers of compute servers and other resources and typically offers their combined capacity on an on-demand, pay-per-cycle basis. The end users of a cloud computing network usually have no idea where the servers are physically located—they just spin up their application and start working

This flexibility is the key advantage to cloud computing, and what distinguishes it from other forms of grid or utility computing and software as a service (SaaS). The ability to launch new instances of an application with minimal labor and expense allows application providers to scale up and down rapidly, recover from a failure, bring up development or test instances, roll out new versions to the customer base

The primary concept behind Cloud Computing isn't a brand new one. John McCarthy within the sixties imagined that processing amenities is going to be supplied to everyone just like a utility. The word "cloud" has already been utilized in numerous contexts such as explaining big ATM systems within the 1990s. Nevertheless, it had been following Google's BOSS Eric Schmidt utilized the term to explain the company type of supplying providers over the Web within 2006. Since then, the term cloud computing has been used mainly as a marketing term in a variety of contexts to represent many different ideas. Certainly, the lack of a standard definition of cloud computing has generated not only market hypes, but also a fair amount of skepticism and confusion. For this reason, recently there has been work on standardizing the definition of cloud computing. As an example, the work in compared over 20 different definitions from a variety of sources to confirm a standard definition. In this paper, we adopt the definition of cloud computing provided by The National Institute of Standards and Technology (NIST), as it covers, in our opinion, all the essential aspects of cloud computing:

**NIST definition of cloud computing[2]** Cloud computing is a model for enabling convenient, on-demand network access to a shared pool of configurable computing resources (e.g., networks, servers, storage, applications, and services) that can be rapidly provisioned and released with minimal management effort or service provider interaction.

The main reason for the existence of different perceptions of cloud computing is that cloud computing, unlike other technical terms, is not a new technology, but rather a new operations model that brings together a set of existing technologies to run business in a different way. Indeed, most of the technologies used by cloud computing, such as virtualization and utility-based pricing, are not new. Instead, cloud computing leverages these existing technologies to meet the technological and economic requirements of today's demand for information technology.

## II. RELATED TECHNOLOGIES

Cloud computing is often compared to the following technologies[3], each of which shares certain aspects with cloud computing:

**Grid Computing**: Grid computing is a distributed computing paradigm that coordinates networked resources to achieve a common computational objective. The development of Grid computing was originally driven by scientific applications which are usually computation-intensive. Cloud computing is similar to Grid computing in that it also employs distributed resources to achieve application-level objectives. However, cloud computing takes one step further by leveraging virtualization technologies at multiple levels (hardware and application platform) to realize resource sharing and dynamic resource provisioning.

**Utility Computing**: Utility computing represents the model of providing resources on-demand and charging customers based on usage rather than a flat rate. Cloud computing can be perceived as a realization of utility computing. It adopts a utility-based pricing scheme entirely for economic reasons. With on-demand resource provisioning and utility based pricing, service providers can truly maximize resource utilization and minimize their operating costs.

**Virtualization**: Virtualization is a technology that abstracts away the details of physical hardware and provides virtualized resources for high-level applications. A virtualized server is commonly called a virtual machine (VM). Virtualization forms the foundation of cloud computing, as it provides the capability of pooling computing resources from clusters of servers and dynamically assigning or reassigning virtual resources to applications on-demand.

**Autonomic Computing**: Originally coined by IBM in 2001, autonomic computing aims at building computing systems capable of self-management, i.e. reacting to internal and external observations without human intervention. The goal of autonomic computing is to overcome the management complexity of today's computer systems. Although cloud computing exhibits certain autonomic features such as automatic resource provisioning, its objective is to lower the resource cost rather than to reduce system complexity. In summary, cloud computing leverages virtualization technology to achieve the goal of providing computing resources as a utility. It shares certain aspects with grid computing and autonomic computing but differs from them in other aspects. Therefore, it offers unique benefits and imposes distinctive challenges to meet its requirements

### III. Hadoop

Hadoop[9] is an open source implementation of the MapReduce parallel processing framework. Hadoop hides the details of parallel processing, including distributing data to processing nodes, restarting subtasks after a failure, and collecting the results of the computation. This framework allows developers to write relatively simple programs that focus on their computation problem, rather than on the nuts and bolts of parallelization.

Hadoop Components

- Distributed file system (HDFS)
    - Single namespace for entire cluster
    - Replicates data 3x for fault-tolerance
- MapReduce framework
    - Executes user jobs specified as "map" and "reduce" functions
    - Manages work distribution & fault-tolerance

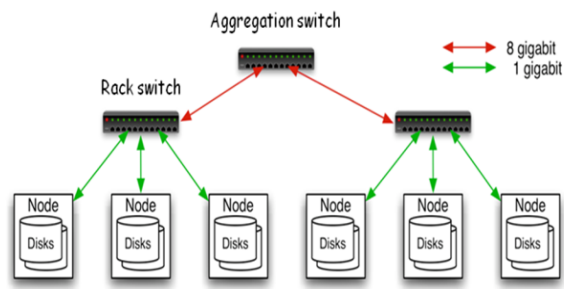

**Fig 1: Typical Hadoop Cluster**

A hadoop Cluster may contain:
- 40 nodes/rack, 1000-4000 nodes in cluster
- 1 Gbps bandwidth within rack, 8 Gbps out of rack
- Node specs (Yahoo terasort): 8 x 2 GHz cores, 8 GB RAM, 4 disks (= 4 TB?)
- Files split into 128MB blocks
- Blocks replicated across several datanodes (usually 3)
- Single namenode stores metadata (file names, block locations, etc)
- Optimized for large files, sequential reads

**a) HDFS- Distributed file system over clouds**

Google File System (GFS) [6] is a proprietary distributed file system developed by Google and specially designed to provide efficient, reliable access to data using large clusters of commodity servers. Files are divided into chunks of 64 megabytes, and are usually appended to or read and only extremely rarely overwritten or shrunk. Compared with traditional file systems, GFS is designed and optimized to run on data centers to provide extremely high data throughputs, low latency and survive individual server failures. Inspired by GFS, the open source Hadoop Distributed File System (HDFS) [4] stores large files across multiple machines. It achieves reliability by replicating the data across multiple servers. Similarly to GFS, data is stored on multiple geo-diverse nodes. The file system is built from a cluster of data nodes, each of which serves blocks of data over the network using a block protocol specific to HDFS. Data is also provided over HTTP, allowing access to all content from a web browser or other types of clients. Data nodes can talk to each other to rebalance data distribution, to move copies around, and to keep the replication of data high.

An advantage of using the HDFS is data awareness between the jobtracker and tasktracker. The jobtracker schedules map/reduce jobs to tasktrackers with an awareness of the data location. An example of this would be if node A contained data (x,y,z) and node B contained data (a,b,c). The jobtracker will schedule node B to perform map/reduce tasks on (a,b,c) and node A would be scheduled to perform map/reduce tasks on (x,y,z). This reduces the amount of traffic that goes over the network and prevents unnecessary data transfer. When Hadoop is used with other filesystems this advantage is not available. This can have a significant impact on the performance of job completion times, which has been demonstrated when running data intensive jobs [10]

### b) Hadoop Mapreduce overview

MapReduce [5] is one of the most popular programming models designed for data centers. It was originally proposed by Google to handle large-scale web search applications and has been proved to be an effective programming model for developing data mining, machine learning and search applications in data centers. In particular, MapReduce can enhance the productivity for junior developers who lack the experience of distributed/parallel development. Hadoop has been successfully used by many companies including AOL, Amazon, Facebook, and New York Times for running their applications on clusters. For example, AOL used it for running an application that analyzes the behavioral pattern of their users so as to offer targeted services.

Although Hadoop is successful in homogeneous computing environments, a performance study conducted by Matei Zaharia et al. [12] shows that MapReduce implemented in the standard distribution of Hadoop is unable to perform well in heterogeneous Cloud computing infrastructure such as Amazon EC2 . Experimental observations reveal that the homogeneity assumptions of MapReduce can cause wrong and often unnecessary speculative execution in heterogeneous environments, sometimes resulting in even worse performance than with speculation disabled. This evaluation and performance results of their enhanced scheduler in Hadoop demonstrate that Cloud execution management systems need to be designed to handle heterogeneity that is present in workloads, applications, and computing infrastructure.
− Commodity machines (cheap, but unreliable)
− Commodity network
− Automatic fault-tolerance (fewer administrators)
− Easy to use (fewer programmers)

As shown in figure.2 a mapper will map the task to a datanode where the data is available. Task trackers will keep track of the work that is being carried by the datanodes.

## IV. PERFORMANCE ISSUES

Several Key factors exist that affect the performance of Hadoop

### a) Cluster Hardware Configuration

Hadoop was designed based on a new approach to storing and processing complex data. Instead of relying on a Storage as Network for massive storage and reliability then moving it to a collection of blades for processing, Hadoop handles large data volumes and reliability in the software tier. Hadoop distributes data across a cluster of balanced machines and uses replication to ensure data reliability and fault tolerance. Because data is distributed on machines with compute power, processing can be sent directly to the machines storing the data. Since each machine in a Hadoop cluster both stores and processes data, they need to be configured to satisfy both data storage and processing requirements. Table:1 gives the summary of the parameters that affect the cluster performance.

| Parameter | Impact / Purpose |
|---|---|
| No.of Cores | Processing Speed |
| RAM | # trips to disk |
| Disks per node | To support rapid scale up |
| Disk speed | High throughput |
| Network Topology | Communication overhead |

Table:1 Hardware Parameters that affect Hadoop Performance

There are four types of nodes in a basic Hadoop cluster. A node referred as a machine performing a particular task. Most of the machines will function as both datanodes and tasktrackers. These nodes both store data and perform processing functions. Recommended specifications for datanodes/tasktrackers in a balanced Hadoop cluster are:

• 4 1TB hard disks in a JBOD (Just a Bunch Of Disks) configuration
• 2 quad core CPUs, running at least 2-2.5GHz

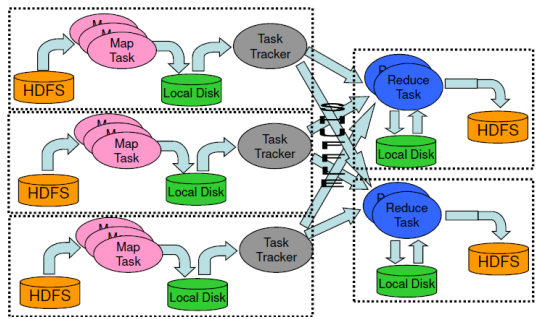

Fig 2: Hadoop Design

- 16-24GBs of RAM (24-32GBs if you're considering HBase)
- Gigabit Ethernet

The namenode is responsible for coordinating data storage on the cluster and the jobtracker for coordinating data processing. The last type of node is the secondarynamenode, which can be colocated on the namenode machine for small clusters, and will run on the same hardware as the namenode for larger clusters. We recommend our customers purchase hardened machines for running the namenodes and jobtrackers, with redundant power and enterprise-grade RAIDed disks. Namenodes also require more RAM relative to the number of data blocks in the cluster. A good rule of thumb is to assume 1GB of namenode memory for every one million blocks stored in the distributed file system. With 100 datanodes in a cluster, 32GBs of RAM on the namenode provides plenty of room to grow. We also recommend having a standby machine to replace the namenode or jobtracker, in the case when one of these fails suddenly.

When you expect your Hadoop cluster to grow beyond 20 machines we recommend that the initial cluster be configured as it were to span two racks, where each rack has a top of rack gigabit switch, and those switches are connected with a 10 GigE interconnect or core switch. Having two logical racks gives the operations team a better understand of the network requirements for inner-rack, and cross-rack communication.

With a Hadoop cluster in place the team can start identifying workloads and prepare to benchmark those workloads to identify CPU and IO bottlenecks. After some time benchmarking and monitoring, the team will have a good understanding as to how additional machines should be configured. It is common to have heterogeneous Hadoop clusters especially as they grow in size. Starting with a set of machines that are not perfect for your workload will not be a waste.

Below is a list of various hardware configurations for different workloads, including our earlier recommendation[10]

- Light Processing Configuration (1U/machine): Two quad core CPUs, 8GB memory, and 4 disk drives (1TB or 2TB). Note that CPU-intensive work such as natural language processing involves loading large models into RAM before processing data and should be configured with 2GB RAM/core instead of 1GB RAM/core.
- Balanced Compute Configuration (1U/machine): Two quad core CPUs, 16 to 24GB memory, and 4 disk drives (1TB or 2TB) directly attached using the motherboard controller. These are often available as twins with two motherboards and 8 drives in a single 2U cabinet.
- Storage Heavy Configuration (2U/machine): Two quad core CPUs, 16 to 24GB memory, and 12 disk drives (1TB or 2TB). The power consumption for this type of machine starts around ~200W in idle state and can go as high as ~350W when active.
- Compute Intensive Configuration (2U/machine): Two quad core CPUs, 48-72GB memory, and 8 disk drives (1TB or 2TB). These are often used when a combination of large in-memory models and heavy reference data caching is required.

Purchasing appropriate hardware for a Hadoop cluster requires benchmarking and careful planning to fully understand the workload. Nevertheless, Hadoop clusters are commonly heterogeneous and we recommend deploying initial hardware with balanced specifications when getting started.

### b) application logic related

#### i) Tune the number of map and reduce tasks appropriately

Tuning the number of map and reduce tasks for a job is important and easy to overlook. Here are some rules of thumb to set these parameters:

- If each task takes less than 30-40 seconds, reduce the number of tasks. The task setup and scheduling overhead is a few seconds, so if tasks finish very quickly, you're wasting time while not doing work. JVM reuse can also be enabled to solve this problem.
- If a job has more than 1TB of input, consider increasing the block size of the input dataset to 256M or even 512M so that the number of tasks will be smaller
- So long as each task runs for at least 30-40 seconds, increase the number of mapper tasks to some multiple of the number of mapper slots in the cluster.
- Don't schedule too many reduce tasks – for most jobs, we recommend a number of reduce tasks equal to or a bit less than the number of reduce slots in the cluster.

#### ii) Take Data locality into consideration

In a cluster where each node has a local disk, it is efficient to move data processing operations to nodes where application data are located. If data are not locally available in a processing node, data have to be migrated via network interconnects to the node that performs the data processing operations. Migrating huge amount of data leads to excessive network congestion, which in turn can deteriorate system performance. HDFS enables Hadoop MapReduce applications to transfer processing operations toward nodes storing application data to be processed by the operations. In a heterogeneous cluster, the computing capacities of nodes may vary significantly. A high-speed node can finish processing data stored in a local disk of the node faster than low-speed counterparts. After a fast node complete the processing of its local input data, the node must support load sharing by handling unprocessed data located in one or more remote slow nodes. When the amount of transferred data due to load sharing is very large, the overhead of moving unprocessed data from slow nodes to fast nodes becomes a critical issue affecting Hadoop's performance. To boost the performance of Hadoop in heterogeneous clusters, we aim to minimize data movement between slow and fast nodes. This goal can be achieved by a data placement scheme[11] that distribute and store data across multiple heterogeneous nodes based on their computing capacities. Data movement can be reduced if the number of file fragments placed on the disk of each node is proportional to the node's data processing speed.To achieve the best I/O performance, one may make replicas of an input data file of a Hadoop application in a way that each node in a Hadoop cluster has a local copy of the input data. Such a data replication scheme can, of course, minimize data transfer among slow and fast nodes in the cluster during the execution of the Hadoop application. The data-replication approach has several limitations. First, it is very expensive to create replicas in a large-scale cluster. Second, distributing a large number of replicas can wasterfully consume scarce network bandwidth in Hadoop clusters. Third, storing replicas requires an unreasonably large amount of disk capacity, which in turn increases the cost of Hadoop clusters. Although all replicas can be produced before the execution of Hadoop applications, significant efforts must be make to reduce the overhead of generating replicas. If the datareplication approach is employed in Hadoop, one has to address the problem of high overhead for creating file replicas by implementing a low-overhead file-replication mechanism. For example, Shen and Zhu developed a proactive lowoverhead file replication scheme for structured peer-to-peer networks [13]. Shen and Zhu's scheme may be incorporated to overcome this limitation.

### c) System Bottlenecks & Resource Under-utilization

#### i) Replication

HDFS is designed to run on highly unreliable hardware. On Yahoo's long-running clusters we observe a node failure rate of 2–3 per 1000 nodes a day. On new (recently out of the factory) nodes, the rate is three times higher. In order to provide data reliability HDFS uses block replication. Initially, each block is replicated by the client to three data-nodes. The block copies are called replicas. A replication factor of three is the default system parameter, which can either be configured or specified per file at creation time.

Once the block is created, its replication is maintained by the system automatically. The name-node detects failed data-nodes, or missing or corrupted individual replicas, and restores their replication by directing the copying of the remaining replicas to other nodes. Replication is the simplest of known data-recovery techniques. Other techniques, such as redundant block striping or erasure codes, are applicable and have been used in other distributed file systems such as GFS, PVFS and Lustre [6, 7, 8]. These approaches, although more space efficient, also involve performance tradeoffs for data recovery. With striping, depending on the redundancy requirements, the system may need to read two or more of the remaining data segments from the nodes it has been striped to in order to reconstruct the missing one. Replication always needs only one copy. For HDFS, the most important advantage of the replication technique is that it provides high availability of data in high demand. This is actively exploited by the MapReduce framework, as it increases replications of configuration and job library files to avoid contention during the job startup, when multiple tasks access the same files simultaneously.

Each block replica on a data-node is represented by a local (native file system) file. The size of this file equals the actual length of the block and does not require extra space to round it up to the maximum block size, as traditional file systems do. Thus, if a block is half full it needs only half of the space of the full block on the local drive. A slight overhead is added, since HDFS also stores a second, smaller metadata file for each block replica, which contains the checksums for the block data.

#### ii) Block reports, heartbeats

The name-node maintains a list of registered data-nodes and blocks belonging to each data-node. A data-node identifies block replicas in its possession to the name-node by sending a block report. A block report contains block ID, length, and the generation stamp for each block replica. The first block report is sent immediately after the data-node registration. It reveals block locations, which are not maintained in the namespace image or in the journal on the name-node. Subsequently, block reports are sent periodically every hour by default and serve as a sanity check, providing that the name-node has an up-to-date view of block replica distribution on the cluster. During normal operation, data-nodes periodically send heartbeats to the name-node to indicate that the data-node is alive. The default heartbeat interval is three seconds. If the name-node does not receive a heartbeat from a data-node in 10 minutes, it pronounces the data-node dead and schedules its blocks for replication on other nodes. Heartbeats also carry information about total and used disk capacity and the number of data transfers currently performed by the node, which plays an important role in the name-node's space and load-balancing decisions. The communication on HDFS clusters is organized in such a way that the name-node does not call data-nodes directly. It uses heartbeats to reply to the data nodes with important instructions. The instructions include commands to:

- Replicate blocks to other nodes. Remove local block replicas
- Re-register or shut down the node
- Send an urgent block report

These commands are important for maintaining the overall system integrity; it is therefore imperative to keep heartbeats frequent even on big clusters. The name-node is optimized to process thousands of heartbeats per second without affecting other name-node operations.

### d) Scale

### i) Namespace Limitations

HDFS is based on an architecture where the namespace is decoupled from the data. The namespace forms the file system metadata, which is maintained by a dedicated server called the *name-node*. The data itself resides on other servers called *data-nodes*. The namespace consists of files and directories. Files are divided into large (128 MB) blocks. To provide data reliability, HDFS uses block replication. Each block by default is replicated to three data-nodes. Once the block is created its replication is maintained by the system automatically. The block copies are called *replicas*.

The name-node keeps the entire namespace in RAM. This architecture has a natural limiting factor: the memory size; that is, the number of namespace objects (files and blocks) the single namespace server can handle. Estimates show that the name-node uses less than 200 bytes to store a single metadata object (a file inode or a block). According to statistics on Y! clusters, a file on average consists of 1.5 blocks. Which means that it takes 600 bytes (1 file object + 2 block objects) to store an average file in name-node's RAM. For example to store 100 million files (referencing 200 million blocks), a name-node should have at least 60 GB of RAM. We have learned by now that the name-node can use 70% of its time to process external client requests. Even with a handful of clients one can saturate the name-node performance by letting the clients send requests to the name-node with very high frequency. The name-node most probably would become unresponsive, potentially sending the whole cluster into a tailspin because internal load requests do not have priority over regular client requests. In practice, the extreme load bursts are uncommon. Regular Hadoop clusters run Map Reduce jobs, and jobs perform conventional file reads or writes. To get or put data from or to HDFS, a client first accesses the name-node and receives block locations, and then directly talks to data-nodes to transfer file data. Thus the frequency of name-node requests is bound by the rate of data transfer from data-nodes

### V. CONCLUSION

In this paper we have presented the overview of Hadoop and several issues that affect the performance of hadoop in heterogeneous clusters in cloud environments. We have also proposed some guidelines on how to overcome these issues to improve the performance of hadoop. As hadoop is open source implementation, we hope our work will provide a better understanding of the performance challenges of Hadoop in heterogeneous clusters, and pave the way for further research in this area.

### REFERENCES

[1] Cloud Computing on Wikipedia, en.wikipedia.org/wiki/Cloudcomputing,
[2] NIST Definition of Cloud Computing v15, csrc.nist.gov/groups/SNS/cloud-computing/cloud-def-v15.doc
[3] Qi Zhang, Lu Cheng, Raouf Boutaba Cloud computing: state-of-the-art and research challenges, J Internet Serv Appl (2010)
[4] Hadoop Distributed File System, hadoop.apache.org/hdfs
[5] HadoopMapReduce, hadoop.apache.org/mapreduce
[6] Ghemawat S, Gobioff H, Leung S-T (2003) The Google file system. In: Proc of SOSP, October 2003
[7] Parallel virtual file system, version 2. http://www.pvfs2.org.
[8] A scalable, high performance file system. http://lustre.org.
[9] http://en.wikipedia.org/wiki/Apache_Hadoop
[10] http://www.cloudera.com/blog/2010/03/clouderas-support-team-shares-some-basic-hardware-recommendations/
[11] M.Zaharia, A.Konwinski, A.Joseph, Y.zatz, and I.Stoica. Improving mapreduce performance in heterogeneous environments. *In OSDI'08: 8th USENIX Symposium on Operating Systems Design and Implementation*, October 2008
[12] J. Dean and S. Ghemawat. Mapreduce: Simplified data processing on large clusters. *OSDI '04*, pages 137–150, 2008.
[13] Haiying Shen and Yingwu Zhu. A proactive low-overhead file replication scheme for structured p2p content delivery networks. *J. Parallel Distrib. Comput.*, 69(5):429–440, 2009.